\documentclass[sigconf]{acmart}
\usepackage{booktabs}
\usepackage{bm}
\usepackage{amsmath}

\AtBeginDocument{%
  \providecommand\BibTeX{{%
    \normalfont B\kern-0.5em{\scshape i\kern-0.25em b}\kern-0.8em\TeX}}}

\copyrightyear{2023}
\acmYear{2023}
\setcopyright{acmlicensed}
\acmConference[CIKM '23] {Proceedings of the 32nd ACM International Conference on Information and Knowledge Management}{October 21--25, 2023}{Birmingham, United Kingdom.}
\acmBooktitle{Proceedings of the 32nd ACM International Conference on Information and Knowledge Management (CIKM '23), October 21--25, 2023, Birmingham, United Kingdom}
\acmPrice{15.00}
\acmISBN{979-8-4007-0124-5/23/10}
\acmDOI{10.1145/XXXXXX.XXXXXX}

\begin{document}

\title{Deep Context Interest Network for Click-Through Rate Prediction}


\author{Xuyang Hou,Zhe Wang,Qi Liu,Tan Qu,Jia Cheng,Jun Lei}
\affiliation{%
  \city{Meituan, Beijing}
  \country{China}}
\email{{houxuyang, wangzhe65, liuqi104, qutan, jia.cheng.sh, leijun}@meituan.com}

\renewcommand{\shortauthors}{}

\begin{abstract}
 Click-Through Rate (CTR) prediction, estimating the probability of a user clicking on an item, is essential in industrial applications, such as online advertising. Many works focus on user behavior modeling to improve CTR prediction performance. However, most of those methods only model users' positive interests from users' click items while ignoring the context information, which is the display items around the clicks, resulting in inferior performance.
In this paper, we highlight the importance of context information on user behavior modeling and propose a novel model named \textbf{D}eep \textbf{C}ontext \textbf{I}nterest \textbf{N}etwork (DCIN), which integrally models the click and its display context to learn users' context-aware interests.
DCIN consists of three key modules: 1) Position-aware Context Aggregation Module (PCAM),  which performs aggregation of display items with an attention mechanism; 2) Feedback-Context Fusion Module (FCFM), which fuses the representation of clicks and display contexts through non-linear feature interaction; 3) Interest Matching Module (IMM), which activates interests related with the target item. 
Moreover, we provide our hands-on solution to implement our DCIN model on large-scale industrial systems.
The significant improvements in both offline and online evaluations demonstrate the superiority of our proposed DCIN method. Notably, DCIN has been deployed on our online advertising system serving the main traffic, which brings 1.5\% CTR and 1.5\% RPM lift.

\end{abstract}



\maketitle

\begin{figure}[htbp]
\begin{center}
\includegraphics[width=0.38\textwidth]{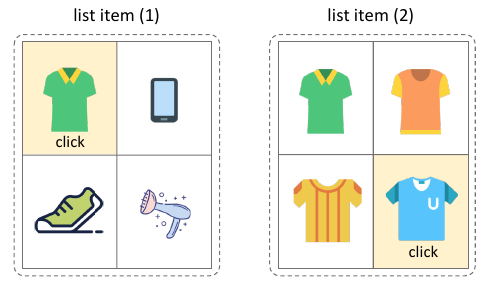}
\end{center}
\caption{The illustration of display context. User's click behavior on an item may be influenced by other display items.}
\label{fig:intro}
\end{figure}

\section{Introduction}
Click-Through Rate (CTR) prediction, which aims at predicting the probability of a user clicking on an item, is critical for industrial applications such as online advertising and recommendation systems~\cite{covington2016deep,xiao2020deep}. Recent research on CTR can be divided into many directions, such as feature interaction and user behavior modeling. This paper focuses on the problem of user behavior modeling.

Works on user behavior modeling hypothesize that users' historical behaviors reflect users' interests towards items. Lots of works~\cite{li2017neural,xu2020deep,zhou2018deep,zhou2019deep,feng2019deep,pi2019practice} focus on capturing user interests from click behaviors and achieve great success in CTR prediction. However, some studies~\cite{pi2019practice,wang2018modeling} find that there are gaps between users' click behaviors and their real interests, leading to biased user representation and inferior performance. In order to tackle the problem, pioneering works DFN~\cite{xie2021deep}, DUMN~\cite{bian2021denoising} introduce users' unclick behaviors to capture users' negative preferences and get further performance promotion. However, they model users' click and unclick behaviors separately, which ignores the interaction between click and unclick items. As shown in Figure~\ref{fig:intro}, there are generally many display items on one screen of a computer/smartphone. Users' click behaviors are not only determined by the matching degree between items and their interests but also influenced by the display context~\cite{zheng2021disentangling,bobadilla2020deepfair,Li_Chen_Dong_Zhang_Wang_Wang_Wang_2023}. For example, as shown in Figure~\ref{fig:intro}, when a user wants to buy a T-shirt, he may click the green one when its surrounding items are smartphones, shoes, or hair dryers since those are boring to him. But when the display items are all T-shirts, he may click the blue one because it may match his interests better. Based on the above analysis, we argue that display context is also essential in interest modeling. RACP~\cite{fan2022modeling} introduces page-wise feedback modeling and aggregates different pages into different interest representations. However, different clicks on one page may indicate different interests, which degrades the model's capability to capture users' fine-grained interests precisely.

In this paper, we propose \textbf{D}eep \textbf{C}ontext \textbf{I}nterest \textbf{N}etwork (DCIN) to extract context-aware interest representations from users' click behaviors and corresponding display contexts aiming at acquiring comprehensive and fine-grained user representation. Specifically, we first design a Position-aware Context Aggregation Module (PCAM), which performs aggregation of display items with an attention mechanism. Meanwhile, we consider position bias~\cite{guo2019pal,huang2021deep} and integrate position information into PCAM. Then, Feedback-Context Fusion Module (FCFM) is proposed to fuse the representation of clicks and corresponding display contexts through non-linear feature interaction to capture users' context-wise interests. Finally, we use Interest Matching Module (IMM) to match behaviors related to the target item. Moreover, we also provide our hands-on solution to implement DCIN on large-scale industrial systems. Our contributions are as follows:
\begin{itemize}
\item We highlight the necessity of introducing context information for interest modeling, which can be used to capture users’ interests more precisely and comprehensively.
\item We devise a novel sequential model, DCIN, which effectively integrates display context and position information to acquire users' context-aware interests.
\item We conduct both offline and online experiments, and the significant improvements demonstrate the superiority of our proposed DCIN method. The online optimization makes DCIN process users' behavior sequences 28x longer than the SOTA method RACP~\cite{fan2022modeling}. Notably, DCIN has served the main traffic in our online advertising system, bringing 1.5\% CTR and 1.5\% RPM lift.
\end{itemize}

\section{METHODOLOGY}
In this section, we introduce the design of DCIN.
\subsection{Preliminaries}
CTR prediction is to predict the probability $p$ that user $u$ clicks target item $t$ with model $\mathcal{F}$, given user behaviors $\bm{C}$ (click items) and $\bm{D}$ (display items), and context $c$, formulated as $p=\mathcal{F}(\bm{C},\bm{D},u,t,c)$.
The click behaviors are a sequential list of click items with corresponding features such as item id, category, etc., i.e., $\bm{C}=\{c_1,c_2\dots,c_N\}$, where $c_i$ is the $i$-th click item and $N$ is the length of $\bm{C}$. To model the display context, we pick out $M$ corresponding display items for each click, hence display behaviors $\bm{D}$ can be denoted as:
\begin{equation}
\label{eq:unclick}
\bm{D}=\{\left[d_{11},d_{12},\dots,d_{1M}\right]\dots,\left[d_{N1},d_{N2},\dots,d_{NM}\right]\}, 
\end{equation}
where $d_{ij}$ is the $j$-th display item around click $c_i$. We try to keep the click in the middle when selecting display items for each click. We denote $p_i$ and $q_{ij}$ as the absolute position of $c_i$ and $d_{ij}$ respectively. We also calculate relative position $r_{ij}=p_i-q_{ij}$ between $c_i$ and $d_{ij}$. We utilize the widely-used embedding technique to transform sparse features into low-dimensional dense vectors. For example, $\bm{e}_{c_i}$ represents the embedding of click item $c_i$.

\subsection{Deep Context Interest Network}
\label{model:DCIN}
\begin{figure*}[htbp]
\begin{center}
\includegraphics[width=0.72\textwidth]{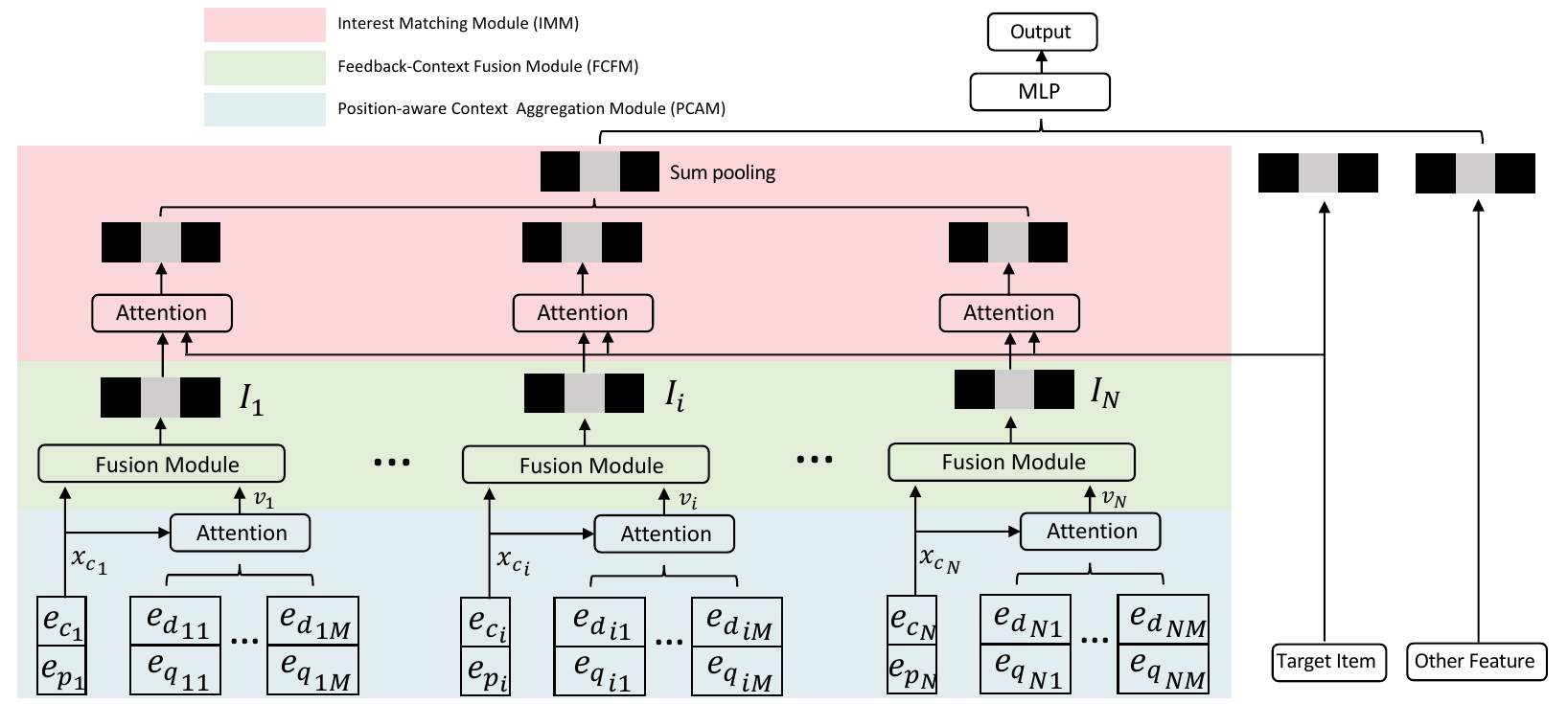}
\end{center}
\caption{The architecture of DCIN network.}
\label{fig:DCIN}
\end{figure*}

The architecture of DCIN is shown in Figure~\ref{fig:DCIN}, mainly consisting of three modules. Specifically, we first design Position-aware Context Aggregation Module (PCAM) to aggregate the display items with an attention mechanism. Then Feedback-Context Fusion Module (FCFM) is proposed to fuse the representation of clicks and corresponding display contexts through non-linear feature interaction to acquire context-aware interests. Finally, we propose the Interest Matching Module (IMM) to learn the interests related to the target item adaptively.

\subsubsection{\textbf{Position-aware Context Aggregation Module}}
As shown in Figure~\ref{fig:intro}, there are generally many display items on one screen of a computer/smartphone. Users' click behaviors are not only determined by the matching degree between items and their interests but also influenced by the display context~\cite{zheng2021disentangling,bobadilla2020deepfair}. However, modeling users' interests only from their click behaviors ignores the context information, leading to biased interest representation and inferior performance. To capture users' context-aware interests, we design PCAM, which aggregates display items for each click with an attention mechanism. Moreover, users generally tend to click top items on the screen (known as position bias~\cite{guo2019pal,huang2021deep}) regardless of the relevance. Hence, items' positions also impact users' click behaviors. We also integrate position information into PCAM for a better understanding of users' actual interests. As a result, PCAM can be formulated as follows:

\begin{equation}
\label{equation:att}
\bm{v}_{i}=\sum_{j=1}^M\mu_{ij}\bm{W}_V\bm{e}_{d_{ij}},
\end{equation}
where $\bm{v_i}$ represents the display context corresponding to click $c_i$ and is modeled as a weighted sum aggregation of display items $\bm{e}_{d_{ij}}$, and $\bm{W}_V$ is learning parameters. $\mu_{ij}$ is the attention weight, which can be formulated as follows:
\begin{equation}
\mu_{ij}=\frac{exp(\alpha_{ij})}{\sum_{j=1}^Mexp(\alpha_{ij})},
\end{equation}
\begin{equation}
\label{att:relavance}
\alpha_{ij}=\bm{W}_Q\bm{x}_{c_i}\cdot \bm{W}_K\bm{x}_{d_{ij}}+e_{r_{ij}},
\end{equation}
where $\alpha_{ij}$ is the relevance between the $i$-th click and its $j$-th display item, $\bm{x}_{c_i}=concat(\bm{e}_{c_i},\bm{e}_{p_i})$ represents the concatenation of item embedding $\bm{e}_{c_i}$ and its absolute position embedding $\bm{e}_{p_i}$ and the same as $\bm{x}_{d_{ij}}$, $e_{r_{ij}}$ is the one-dimensional embedding vector of relative position $p_{ij}$, and $\bm{W}_Q$ and $\bm{W}_K$ are learning parameters.

\subsubsection{\textbf{Feedback-Context Fusion Module}} Previous user behavior models~\cite{li2017neural,xu2020deep,zhou2018deep,zhou2019deep,feng2019deep,pi2019practice} directly regard users' click items as users' interest, ignoring the context information. To capture users' context-aware interests, we propose FCFM to fuse the representation of clicks and corresponding display contexts through non-linear feature interaction. Concretely, FCFM is implemented by a 2-layer MLP:
\begin{equation}
    \bm{f}^{(k+1)}_i=ReLU(\bm{W}^{(k)}\bm{f}^{(k)}+\bm{b}^{(k)}),
\end{equation}
where $\bm{f}^{(k)}_i$ is the output of the $k$-th layer, and $\bm{W}^{(k)}$ and $\bm{b}^{(k)}$ are learning parameters. The input is formulated as follows:
\begin{equation}
    \bm{f}^{(0)}_i=concat(\bm{x}_{c_i}, \bm{v}_{i}, \bm{x}_{c_i}-\bm{v}_{i}, \bm{x}_{c_i}\ast \bm{v}_{i}),
\end{equation}
where $\ast$ represents the element-wise product. Then, we obtain context-aware interest $\bm{I}_i = \bm{f}^2_i$ from click $c_i$ and its corresponding display context $\bm{v}_i$. For generalization and trainability, we share the learnable parameters across different clicks. 

\subsubsection{\textbf{Interest Matching Module}}
With the first two modules, we integrally model the click and its display context to capture users’ context-aware interests, denoted as $\bm{I}=\{\bm{I}_1, \bm{I}_2, \dots, \bm{I}_N\}$, which are more precise and comprehensive than those extracted only from click behaviors. To capture the user interests that are related to the target item, the SOTA sequential models can be easily integrated, like DIN~\cite{zhou2018deep}, DIEN~\cite{zhou2019deep}, BST~\cite{chen2019behavior}, etc. Here we choose the DIN method to aggregate user interests, which can be formulated as follows:
\begin{equation}
    \bm{U}=\sum_{i=1}^N\frac{exp(\bm{e}_t\cdot\bm{I}_i)}{\sum_{j=1}^Nexp(\bm{e}_t\cdot\bm{I}_j)}\cdot\bm{I}_i,
\end{equation}
where $\bm{U}$ is the final user representation, which will be concatenated with other features and fed into MLP for CTR prediction. We utilize the widely-used negative log-likelihood function as the loss function to train our DCIN model.

\subsection{Optimization for Online Serving}
\label{online opt}
For industrial recommender systems, the serving latency constraints are highly restricted. Since we introduce display items to capture users' context-aware interests, the sequence length has been extended tens of times. Generally, the serving latency of online models is proportional to the sequence length. As a result, it is challenging to make DCIN serve in an industrial online system directly. To address the challenge, we make some online optimizations. Specifically, since PCAM and FCFM have no connection with the target item, we can pre-calculate users' context-aware interest representations $\bm{I}=\{\bm{I}_1, \bm{I}_2, \dots, \bm{I}_N\}$, and directly make it as model's input when online serving, which saves a large amount of online serving time (10 milliseconds). Since RACP~\cite{fan2022modeling} needs the target item to aggregate page-wise interests, it can only shorten the sequence length to make the model serve online. With online optimization, DCIN can process users' behavior sequences 28x longer than the SOTA method RACP, adding only one millisecond of online latency.

\section{Experiments}
\subsection{Experimental Settings}
\subsubsection{\textbf{Datasets}}
Integrating display context and position information into interest models is a novel exploration. Hence we build a 31-day industrial dataset from our online advertising system, which contains more than 1 billion samples. Following the standard industry practice, we split the data in the first 30 days as the training set and the data on the last day as the testing set. Users' recent 50-click behaviors are reserved. The surrounding 20 display items for each click are utilized to model the display context.
\subsubsection{\textbf{Compared Methods}}
We implement six state-of-art models to compare with DCIN. All models use the same features for fair comparisons.

\textbf{Wide\&Deep}~\cite{cheng2016wide} consists of a Wide part and a Deep part to learn raw and high-order feature interactions.

\textbf{DeepFM}~\cite{guo2017deepfm} replaces the Wide part in~\cite{cheng2016wide} with an FM layer.

\textbf{DIN}~\cite{zhou2018deep} introduces an attention mechanism to capture users' diverse interests for different target items.

\textbf{DIEN}~\cite{zhou2019deep} proposes an interest evolving layer based on a gated recurrent unit (GRU) to model users' dynamic interests.

\textbf{DFN}~\cite{xie2021deep} tries to learn positive preferences from click behaviors and negative preferences from unclick behaviors. 

\textbf{RACP}~\cite{fan2022modeling} introduces page-wise feedback modeling and is the published SOTA solution for user behavior modeling.

\subsubsection{\textbf{Evaluation Metrics}}
We utilize a widely-adopted metric Area Under Curve (AUC) for evaluation in the CTR prediction task. We further introduce RelaImpr~\cite{yan2014coupled} to measure the relative improvements over the base model:
\begin{equation}
    RelaImpr=\frac{AUC(measured\ model)-0.5}{AUC(base\ model)-0.5}-1
\end{equation}
\subsection{Experimental Results}
We repeat all experiments 5 times and report their average results. In experiments, the embedding dimension is 16 for each feature. The hidden sizes of MLP in FCFM and the output layer are 64/16, and 1024/512/128 respectively. We utilize Adam with the learning rate to be 1e-4, and the batch size is 3000.
\subsubsection{\textbf{Offline Performance Comparison}}
The experimental results are shown in Table~\ref{tab:perform_auc}, from which we can find that our proposed DCIN significantly outperforms all methods on AUC and achieves 21.24\% RelaImpr over the base model. As introduced in Section~\ref{online opt}, DCIN can process users' behavior sequences 28x longer than RACP. For a fair comparison, DCIN (short) uses the same sequence length as RACP, and DCIN uses a longer sequence as implemented in our online advertising system. Specifically, first, DIN and DIEN perform much better than Wide\&Deep and DeepFM because they capture users' diverse and temporal interests for different target items. Second, DFN beats DIN and DIEN slightly since it tries to capture dislike signals from unclick behaviors. Fourth, RACP performs better than DFN since it models users' contextualized page-wise interests. Fifth, our proposed DCIN (short) beats RACP. On the one hand, different clicks, even on one page, may indicate different interests. RACP only models page-wise interests, degrading the model's capability to capture users' fine-grained interests. On the other hand, RACP ignores the impact that items' positions have on users' click behaviors. Finally, DCIN beats DCIN (short) since it introduces a longer sequence.

\begin{table}[htbp]
\centering
\caption{Performance comparison with different methods.}
\label{tab:perform_auc}
\begin{tabular}{ccccc}
\toprule
Model & AUC & RelaImpr \\
\midrule
Wide\&Deep & 0.63588 & 0.00\% \\
DeepFM & 0.63602 & 0.10\% \\
\midrule
DIN & 0.65012 & 10.48\% \\
DIEN & 0.65102 & 11.14\% \\
\midrule
DFN & 0.65220 & 12.01\% \\
\midrule
RACP & 0.65872 & 16.81\% \\
DCIN (short) & 0.66186 & 19.12\% \\
\textbf{DCIN*} & \textbf{0.66475} & \textbf{21.24\%} \\
\midrule
DCIN (short)-w/o position & 0.65850 & 16.64\% \\
DCIN (short)-w/o FCFM & 0.65950 & 17.38\% \\
\bottomrule
\end{tabular}
\end{table}

\subsubsection{\textbf{Ablation Tests}}
We also conduct ablation tests to verify the effectiveness and necessity of different components in DCIN. Firstly, we remove the position information in PCAM and conduct experiment DCIN (short)-w/o position. As shown in Table~\ref{tab:perform_auc}, DCIN (short)-w/o position performs worse than DCIN (short) since items' positions also have important impact on users' click behaviors. Secondly, we remove the FCFM module in DCIN and conduct experiment DCIN (short)-w/o FCFM where we directly regard display context $\bm{v}_i$ as users' interests, which can be formulated as $\bm{I}_i = \bm{v}_i$ instead of $\bm{I}_i = \bm{f}^2_i$. As shown in Table~\ref{tab:perform_auc}, DCIN (short)-w/o FCFM also performs poorly than DCIN (short) due to the lack of fusion between clicks and display contexts.

\subsubsection{\textbf{Online A/B Test}}
We conduct a strict online A/B test to evaluate DCIN on our online advertising system for two weeks. Compared with DIN (online baseline), DCIN has improved CTR by 1.5\% and RPM (Revenue Per Mile) by 1.5\%. For now, DCIN has been deployed online and serves the main traffic, which contributes to significant business revenue growth.

\subsubsection{\textbf{Case Study}}
\begin{figure}[htbp]
\begin{center}
\includegraphics[width=0.35\textwidth]{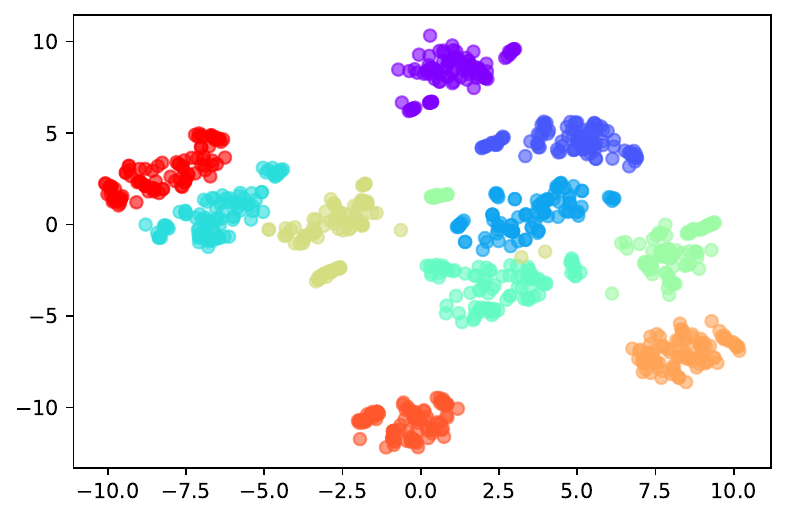}
\end{center}
\caption{Visualization of context-aware interests for 10 click items under 100 different display contexts. Each color represents one click item. The representation distribution shows the separation among click items and the diversity among different display contexts of the same click item.}
\label{fig:tsne}
\end{figure}

To demonstrate the effectiveness of DCIN for context-aware interest representations, we randomly select 10 click items and choose 100 different display contexts for each click respectively. Then we extract the corresponding context-aware interests $\bm{I}_i$ and visualize their distribution through t-SNE~\cite{van2008visualizing}. As shown in Figure~\ref{fig:tsne}, the distinct separation among click items indicates the dominance of click behavior in learning. The obvious diversity of different contexts of the same click demonstrates the ability of DCIN to capture fine-grained context-aware interest representations.

\begin{figure}[htbp]
\begin{center}
\includegraphics[width=0.37\textwidth]{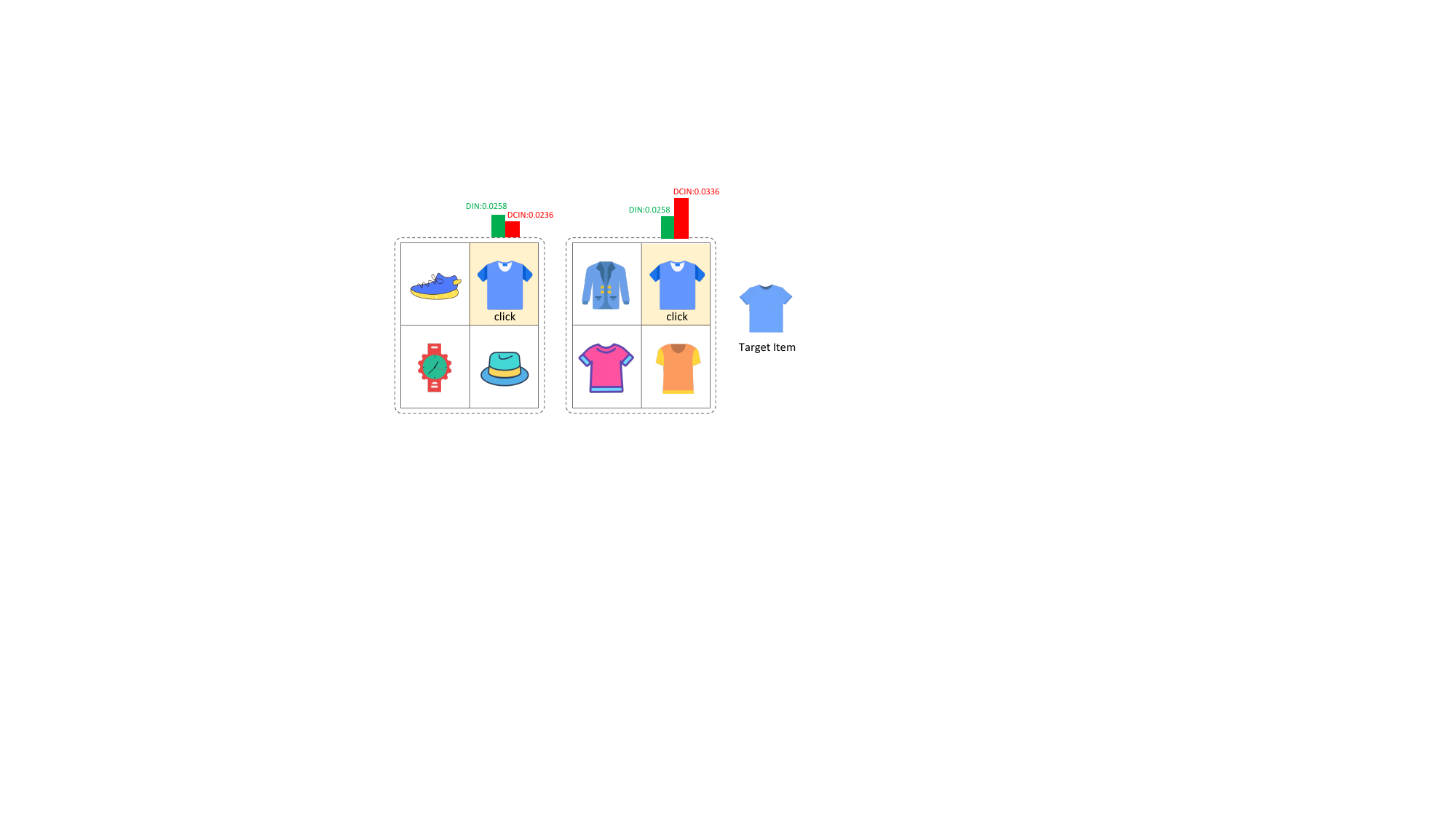}
\end{center}
\caption{Case study of attention weights between the target item and the same click item under different display context. DIN produces the same attention weights due to missing context information. The proposed DCIN can assign higher weight to the click under competitive display context.}
\label{fig:weights}
\end{figure}

To verify the effectiveness of context modeling, we visualize the attention weights from IMM between the target item and the same click item under different display contexts. As shown in Figure~\ref{fig:weights}, the weights of DIN (green color) is the same since DIN only takes click behaviors as input, resulting in the inability to distinguish different display contexts. While DCIN learns higher weights under the right competitive display context where all display items are of interest to the user.

\section{Conclusions}
In this paper, we highlight the necessity of display contexts and position information for interest modeling and devise a novel sequential model DCIN. The significant improvements in both offline and online evaluations demonstrate the superiority of our proposed DCIN methods. Notably, DCIN has been deployed on our online advertising system, and now it is serving the main traffic. 
\bibliographystyle{ACM-Reference-Format}
\balance 
\bibliography{sample-base}


\begin{thebibliography}{22}


\ifx \showCODEN    \undefined \def \showCODEN     #1{\unskip}     \fi
\ifx \showDOI      \undefined \def \showDOI       #1{#1}\fi
\ifx \showISBNx    \undefined \def \showISBNx     #1{\unskip}     \fi
\ifx \showISBNxiii \undefined \def \showISBNxiii  #1{\unskip}     \fi
\ifx \showISSN     \undefined \def \showISSN      #1{\unskip}     \fi
\ifx \showLCCN     \undefined \def \showLCCN      #1{\unskip}     \fi
\ifx \shownote     \undefined \def \shownote      #1{#1}          \fi
\ifx \showarticletitle \undefined \def \showarticletitle #1{#1}   \fi
\ifx \showURL      \undefined \def \showURL       {\relax}        \fi
\providecommand\bibfield[2]{#2}
\providecommand\bibinfo[2]{#2}
\providecommand\natexlab[1]{#1}
\providecommand\showeprint[2][]{arXiv:#2}

\bibitem[Bian et~al\mbox{.}(2021)]%
        {bian2021denoising}
\bibfield{author}{\bibinfo{person}{Zhi Bian}, \bibinfo{person}{Shaojun Zhou},
  \bibinfo{person}{Hao Fu}, \bibinfo{person}{Qihong Yang},
  \bibinfo{person}{Zhenqi Sun}, \bibinfo{person}{Junjie Tang},
  \bibinfo{person}{Guiquan Liu}, \bibinfo{person}{Kaikui Liu}, {and}
  \bibinfo{person}{Xiaolong Li}.} \bibinfo{year}{2021}\natexlab{}.
\newblock \showarticletitle{Denoising User-aware Memory Network for
  Recommendation}. In \bibinfo{booktitle}{\emph{Fifteenth ACM Conference on
  Recommender Systems}}. \bibinfo{pages}{400--410}.
\newblock


\bibitem[Bobadilla et~al\mbox{.}(2020)]%
        {bobadilla2020deepfair}
\bibfield{author}{\bibinfo{person}{Jes{\'u}s Bobadilla},
  \bibinfo{person}{Ra{\'u}l Lara-Cabrera}, \bibinfo{person}{{\'A}ngel
  Gonz{\'a}lez-Prieto}, {and} \bibinfo{person}{Fernando Ortega}.}
  \bibinfo{year}{2020}\natexlab{}.
\newblock \showarticletitle{Deepfair: deep learning for improving fairness in
  recommender systems}.
\newblock \bibinfo{journal}{\emph{arXiv preprint arXiv:2006.05255}}
  (\bibinfo{year}{2020}).
\newblock


\bibitem[Chen et~al\mbox{.}(2019)]%
        {chen2019behavior}
\bibfield{author}{\bibinfo{person}{Qiwei Chen}, \bibinfo{person}{Huan Zhao},
  \bibinfo{person}{Wei Li}, \bibinfo{person}{Pipei Huang}, {and}
  \bibinfo{person}{Wenwu Ou}.} \bibinfo{year}{2019}\natexlab{}.
\newblock \showarticletitle{Behavior sequence transformer for e-commerce
  recommendation in alibaba}. In \bibinfo{booktitle}{\emph{Proceedings of the
  1st International Workshop on Deep Learning Practice for High-Dimensional
  Sparse Data}}. \bibinfo{pages}{1--4}.
\newblock


\bibitem[Cheng et~al\mbox{.}(2016)]%
        {cheng2016wide}
\bibfield{author}{\bibinfo{person}{Heng-Tze Cheng}, \bibinfo{person}{Levent
  Koc}, \bibinfo{person}{Jeremiah Harmsen}, \bibinfo{person}{Tal Shaked},
  \bibinfo{person}{Tushar Chandra}, \bibinfo{person}{Hrishi Aradhye},
  \bibinfo{person}{Glen Anderson}, \bibinfo{person}{Greg Corrado},
  \bibinfo{person}{Wei Chai}, \bibinfo{person}{Mustafa Ispir}, {et~al\mbox{.}}}
  \bibinfo{year}{2016}\natexlab{}.
\newblock \showarticletitle{Wide \& deep learning for recommender systems}. In
  \bibinfo{booktitle}{\emph{Proceedings of the 1st workshop on deep learning
  for recommender systems}}. \bibinfo{pages}{7--10}.
\newblock


\bibitem[Covington et~al\mbox{.}(2016)]%
        {covington2016deep}
\bibfield{author}{\bibinfo{person}{Paul Covington}, \bibinfo{person}{Jay
  Adams}, {and} \bibinfo{person}{Emre Sargin}.}
  \bibinfo{year}{2016}\natexlab{}.
\newblock \showarticletitle{Deep neural networks for youtube recommendations}.
  In \bibinfo{booktitle}{\emph{Proceedings of the 10th ACM conference on
  recommender systems}}. \bibinfo{pages}{191--198}.
\newblock


\bibitem[Fan et~al\mbox{.}(2022)]%
        {fan2022modeling}
\bibfield{author}{\bibinfo{person}{Zhifang Fan}, \bibinfo{person}{Dan Ou},
  \bibinfo{person}{Yulong Gu}, \bibinfo{person}{Bairan Fu},
  \bibinfo{person}{Xiang Li}, \bibinfo{person}{Wentian Bao},
  \bibinfo{person}{Xin-Yu Dai}, \bibinfo{person}{Xiaoyi Zeng},
  \bibinfo{person}{Tao Zhuang}, {and} \bibinfo{person}{Qingwen Liu}.}
  \bibinfo{year}{2022}\natexlab{}.
\newblock \showarticletitle{Modeling Users’ Contextualized Page-wise Feedback
  for Click-Through Rate Prediction in E-commerce Search}.
\newblock  (\bibinfo{year}{2022}).
\newblock


\bibitem[Feng et~al\mbox{.}(2019)]%
        {feng2019deep}
\bibfield{author}{\bibinfo{person}{Yufei Feng}, \bibinfo{person}{Fuyu Lv},
  \bibinfo{person}{Weichen Shen}, \bibinfo{person}{Menghan Wang},
  \bibinfo{person}{Fei Sun}, \bibinfo{person}{Yu Zhu}, {and}
  \bibinfo{person}{Keping Yang}.} \bibinfo{year}{2019}\natexlab{}.
\newblock \showarticletitle{Deep session interest network for click-through
  rate prediction}.
\newblock \bibinfo{journal}{\emph{arXiv preprint arXiv:1905.06482}}
  (\bibinfo{year}{2019}).
\newblock


\bibitem[Guo et~al\mbox{.}(2017)]%
        {guo2017deepfm}
\bibfield{author}{\bibinfo{person}{Huifeng Guo}, \bibinfo{person}{Ruiming
  Tang}, \bibinfo{person}{Yunming Ye}, \bibinfo{person}{Zhenguo Li}, {and}
  \bibinfo{person}{Xiuqiang He}.} \bibinfo{year}{2017}\natexlab{}.
\newblock \showarticletitle{DeepFM: a factorization-machine based neural
  network for CTR prediction}.
\newblock \bibinfo{journal}{\emph{arXiv preprint arXiv:1703.04247}}
  (\bibinfo{year}{2017}).
\newblock


\bibitem[Guo et~al\mbox{.}(2019)]%
        {guo2019pal}
\bibfield{author}{\bibinfo{person}{Huifeng Guo}, \bibinfo{person}{Jinkai Yu},
  \bibinfo{person}{Qing Liu}, \bibinfo{person}{Ruiming Tang}, {and}
  \bibinfo{person}{Yuzhou Zhang}.} \bibinfo{year}{2019}\natexlab{}.
\newblock \showarticletitle{PAL: a position-bias aware learning framework for
  CTR prediction in live recommender systems}. In
  \bibinfo{booktitle}{\emph{Proceedings of the 13th ACM Conference on
  Recommender Systems}}. \bibinfo{pages}{452--456}.
\newblock


\bibitem[Huang et~al\mbox{.}(2021)]%
        {huang2021deep}
\bibfield{author}{\bibinfo{person}{Jianqiang Huang}, \bibinfo{person}{Ke Hu},
  \bibinfo{person}{Qingtao Tang}, \bibinfo{person}{Mingjian Chen},
  \bibinfo{person}{Yi Qi}, \bibinfo{person}{Jia Cheng}, {and}
  \bibinfo{person}{Jun Lei}.} \bibinfo{year}{2021}\natexlab{}.
\newblock \showarticletitle{Deep Position-wise Interaction Network for CTR
  Prediction}. In \bibinfo{booktitle}{\emph{Proceedings of the 44th
  International ACM SIGIR Conference on Research and Development in Information
  Retrieval}}. \bibinfo{pages}{1885--1889}.
\newblock


\bibitem[Li et~al\mbox{.}(2017)]%
        {li2017neural}
\bibfield{author}{\bibinfo{person}{Jing Li}, \bibinfo{person}{Pengjie Ren},
  \bibinfo{person}{Zhumin Chen}, \bibinfo{person}{Zhaochun Ren},
  \bibinfo{person}{Tao Lian}, {and} \bibinfo{person}{Jun Ma}.}
  \bibinfo{year}{2017}\natexlab{}.
\newblock \showarticletitle{Neural attentive session-based recommendation}. In
  \bibinfo{booktitle}{\emph{Proceedings of the 2017 ACM on Conference on
  Information and Knowledge Management}}. \bibinfo{pages}{1419--1428}.
\newblock


\bibitem[Li et~al\mbox{.}(2023)]%
        {Li_Chen_Dong_Zhang_Wang_Wang_Wang_2023}
\bibfield{author}{\bibinfo{person}{Xiang Li}, \bibinfo{person}{Shuwei Chen},
  \bibinfo{person}{Jian Dong}, \bibinfo{person}{Jin Zhang},
  \bibinfo{person}{Yongkang Wang}, \bibinfo{person}{Xingxing Wang}, {and}
  \bibinfo{person}{Dong Wang}.} \bibinfo{year}{2023}\natexlab{}.
\newblock \showarticletitle{Decision-Making Context Interaction Network for
  Click-Through Rate Prediction}.
\newblock \bibinfo{journal}{\emph{Proceedings of the AAAI Conference on
  Artificial Intelligence}} \bibinfo{volume}{37}, \bibinfo{number}{4}
  (\bibinfo{date}{Jun.} \bibinfo{year}{2023}), \bibinfo{pages}{5195--5202}.
\newblock


\bibitem[Pi et~al\mbox{.}(2019)]%
        {pi2019practice}
\bibfield{author}{\bibinfo{person}{Qi Pi}, \bibinfo{person}{Weijie Bian},
  \bibinfo{person}{Guorui Zhou}, \bibinfo{person}{Xiaoqiang Zhu}, {and}
  \bibinfo{person}{Kun Gai}.} \bibinfo{year}{2019}\natexlab{}.
\newblock \showarticletitle{Practice on long sequential user behavior modeling
  for click-through rate prediction}. In \bibinfo{booktitle}{\emph{Proceedings
  of the 25th ACM SIGKDD International Conference on Knowledge Discovery \&
  Data Mining}}. \bibinfo{pages}{2671--2679}.
\newblock


\bibitem[Van~der Maaten and Hinton(2008)]%
        {van2008visualizing}
\bibfield{author}{\bibinfo{person}{Laurens Van~der Maaten} {and}
  \bibinfo{person}{Geoffrey Hinton}.} \bibinfo{year}{2008}\natexlab{}.
\newblock \showarticletitle{Visualizing data using t-SNE.}
\newblock \bibinfo{journal}{\emph{Journal of machine learning research}}
  \bibinfo{volume}{9}, \bibinfo{number}{11} (\bibinfo{year}{2008}).
\newblock


\bibitem[Wang et~al\mbox{.}(2018)]%
        {wang2018modeling}
\bibfield{author}{\bibinfo{person}{Menghan Wang}, \bibinfo{person}{Mingming
  Gong}, \bibinfo{person}{Xiaolin Zheng}, {and} \bibinfo{person}{Kun Zhang}.}
  \bibinfo{year}{2018}\natexlab{}.
\newblock \showarticletitle{Modeling dynamic missingness of implicit feedback
  for recommendation}.
\newblock \bibinfo{journal}{\emph{Advances in neural information processing
  systems}}  \bibinfo{volume}{31} (\bibinfo{year}{2018}).
\newblock


\bibitem[Xiao et~al\mbox{.}(2020)]%
        {xiao2020deep}
\bibfield{author}{\bibinfo{person}{Zhibo Xiao}, \bibinfo{person}{Luwei Yang},
  \bibinfo{person}{Wen Jiang}, \bibinfo{person}{Yi Wei}, \bibinfo{person}{Yi
  Hu}, {and} \bibinfo{person}{Hao Wang}.} \bibinfo{year}{2020}\natexlab{}.
\newblock \showarticletitle{Deep multi-interest network for click-through rate
  prediction}. In \bibinfo{booktitle}{\emph{Proceedings of the 29th ACM
  International Conference on Information \& Knowledge Management}}.
  \bibinfo{pages}{2265--2268}.
\newblock


\bibitem[Xie et~al\mbox{.}(2021)]%
        {xie2021deep}
\bibfield{author}{\bibinfo{person}{Ruobing Xie}, \bibinfo{person}{Cheng Ling},
  \bibinfo{person}{Yalong Wang}, \bibinfo{person}{Rui Wang},
  \bibinfo{person}{Feng Xia}, {and} \bibinfo{person}{Leyu Lin}.}
  \bibinfo{year}{2021}\natexlab{}.
\newblock \showarticletitle{Deep feedback network for recommendation}. In
  \bibinfo{booktitle}{\emph{Proceedings of the Twenty-Ninth International
  Conference on International Joint Conferences on Artificial Intelligence}}.
  \bibinfo{pages}{2519--2525}.
\newblock


\bibitem[Xu et~al\mbox{.}(2020)]%
        {xu2020deep}
\bibfield{author}{\bibinfo{person}{Weinan Xu}, \bibinfo{person}{Hengxu He},
  \bibinfo{person}{Minshi Tan}, \bibinfo{person}{Yunming Li},
  \bibinfo{person}{Jun Lang}, {and} \bibinfo{person}{Dongbai Guo}.}
  \bibinfo{year}{2020}\natexlab{}.
\newblock \showarticletitle{Deep interest with hierarchical attention network
  for click-through rate prediction}. In \bibinfo{booktitle}{\emph{Proceedings
  of the 43rd International ACM SIGIR Conference on Research and Development in
  Information Retrieval}}. \bibinfo{pages}{1905--1908}.
\newblock


\bibitem[Yan et~al\mbox{.}(2014)]%
        {yan2014coupled}
\bibfield{author}{\bibinfo{person}{Ling Yan}, \bibinfo{person}{Wu-Jun Li},
  \bibinfo{person}{Gui-Rong Xue}, {and} \bibinfo{person}{Dingyi Han}.}
  \bibinfo{year}{2014}\natexlab{}.
\newblock \showarticletitle{Coupled group lasso for web-scale ctr prediction in
  display advertising}. In \bibinfo{booktitle}{\emph{International Conference
  on Machine Learning}}. PMLR, \bibinfo{pages}{802--810}.
\newblock


\bibitem[Zheng et~al\mbox{.}(2021)]%
        {zheng2021disentangling}
\bibfield{author}{\bibinfo{person}{Yu Zheng}, \bibinfo{person}{Chen Gao},
  \bibinfo{person}{Xiang Li}, \bibinfo{person}{Xiangnan He},
  \bibinfo{person}{Yong Li}, {and} \bibinfo{person}{Depeng Jin}.}
  \bibinfo{year}{2021}\natexlab{}.
\newblock \showarticletitle{Disentangling user interest and conformity for
  recommendation with causal embedding}. In
  \bibinfo{booktitle}{\emph{Proceedings of the Web Conference 2021}}.
  \bibinfo{pages}{2980--2991}.
\newblock


\bibitem[Zhou et~al\mbox{.}(2019)]%
        {zhou2019deep}
\bibfield{author}{\bibinfo{person}{Guorui Zhou}, \bibinfo{person}{Na Mou},
  \bibinfo{person}{Ying Fan}, \bibinfo{person}{Qi Pi}, \bibinfo{person}{Weijie
  Bian}, \bibinfo{person}{Chang Zhou}, \bibinfo{person}{Xiaoqiang Zhu}, {and}
  \bibinfo{person}{Kun Gai}.} \bibinfo{year}{2019}\natexlab{}.
\newblock \showarticletitle{Deep interest evolution network for click-through
  rate prediction}. In \bibinfo{booktitle}{\emph{Proceedings of the AAAI
  conference on artificial intelligence}}, Vol.~\bibinfo{volume}{33}.
  \bibinfo{pages}{5941--5948}.
\newblock


\bibitem[Zhou et~al\mbox{.}(2018)]%
        {zhou2018deep}
\bibfield{author}{\bibinfo{person}{Guorui Zhou}, \bibinfo{person}{Xiaoqiang
  Zhu}, \bibinfo{person}{Chenru Song}, \bibinfo{person}{Ying Fan},
  \bibinfo{person}{Han Zhu}, \bibinfo{person}{Xiao Ma},
  \bibinfo{person}{Yanghui Yan}, \bibinfo{person}{Junqi Jin},
  \bibinfo{person}{Han Li}, {and} \bibinfo{person}{Kun Gai}.}
  \bibinfo{year}{2018}\natexlab{}.
\newblock \showarticletitle{Deep interest network for click-through rate
  prediction}. In \bibinfo{booktitle}{\emph{Proceedings of the 24th ACM SIGKDD
  international conference on knowledge discovery \& data mining}}.
  \bibinfo{pages}{1059--1068}.
\newblock


\end{thebibliography}

\appendix

\end{document}